\documentclass[a4paper,fleqn]{cas-sc}
\usepackage[authoryear,longnamesfirst]{natbib}
\usepackage{amsthm}
\usepackage{todonotes} 
\usepackage{graphicx}
\usepackage{gensymb}
\usepackage{textcomp}
\usepackage{ulem}
\usepackage{caption}
\usepackage{lineno}
\ExplSyntaxOn
\keys_set:nn { stm / mktitle } { nologo }
\ExplSyntaxOff

\graphicspath{ {./images/} }

\def\tsc#1{\csdef{#1}{\textsc{\lowercase{#1}}\xspace}}
\tsc{WGM}
\tsc{QE}
\tsc{EP}
\tsc{PMS}
\tsc{BEC}
\tsc{DE}

\newcommand*{\figref}[2][]{%
  \hyperref[{fig:#2}]{%
    Fig.~\ref*{fig:#2}%
    \ifx\\#1\\%
    \else
      .#1%
    \fi
  }%
}

\newproof{pf}{Proof}

\begin{document}
\let\WriteBookmarks\relax
\def\floatpagepagefraction{1}
\def\textpagefraction{.001}

\shorttitle{COTS-Capsule Spaceborne Payload} 

\shortauthors{Simhony Yoav et~al.}

\title [mode = title]{Spaceborne COTS-Capsule Hodoscope: Detecting and Characterizing Particle Radiation}

\author[1,2]{Simhony Yoav}[type=editor]
\ead{yoavsimhony@mail.tau.ac.il}


\affiliation[1]{organization={Tel Aviv University},
    addressline={Raymond and Beverly Sackler School of Physics and Astronomy}, 
    city={Tel Aviv},
    postcode={69978}, 
    country={Israel}}
    
    \affiliation[2]{organization={Tel Aviv University},
    addressline={School of Electrical Engineering}, 
    city={Tel aviv},
    postcode={69978}, 
    country={Israel}}

\author%
[3]
{Segal Alex}
\ead{alexs@afeka.ac.il}

\author [1] {Orlov Yuri}[]
\ead{ureor1@gmail.com}

\author[1,2]{Bashi Dolev}[]
\ead{dolevbashi@gmail.com}


\author%
[2]
{Amrani Ofer}
\ead{ofera@tauex.tau.ac.il}

\author%
[1,4]
{Etzion Erez}
\ead{ereze@tauex.tau.ac.il}

\affiliation[3]{organization={Afeka College of Engineering},
    addressline={Unit of Mathematics}, 
    city={Tel Aviv},
    postcode={6910721}, 
    country={Israel}}

\affiliation[4]{organization={University of Toronto},
    addressline={Department of Physics}, 
    city={Toronto},
    postcode={M5S 1A2}, 
    state={ON},
    country={Canada}}    

\begin{abstract}
The COTS-Capsule Spaceborne hodoscope was launched into low earth orbit aboard the International Space Station and operated during 2021-2022. The primary objectives of the payload are: measuring and characterizing the radiation environment within the space station; serving as a technology demonstrator for the COTS-Capsule radiation mitigation apparatus; and testing the interaction of high-energy cosmic particles in space with our detectors. The payload features a particle hodoscope equipped with an array of novel detectors based on polyvinyl toluene scintillators and silicon photomultiplier sensors that are used for radiation detection and characterization. This paper provides a comprehensive account of the construction, preflight performance, testing, qualification, and in-orbit calibration of the COTS-Capsule payload. The hodoscope, sensitive to ionizing cosmic particles, facilitates impinging particle track reconstruction, energy deposition estimation, and linear energy transfer estimation. In specific instances, it also provides particle identification and energy deposition estimation.
\end{abstract}

\begin{keywords}
Space \sep Cosmic particles \sep Particle detector \sep International Space Station (ISS) \sep GEANT4 \sep Hodoscope
\end{keywords}

\maketitle

\section{Introduction}

Primary cosmic radiation consists of high-energy particles from astrophysical sources throughout the universe. These particles continuously impact artificial satellites and spacecraft, causing disruptions to onboard electronics, optics, and biological tissue. Traveling at velocities approaching the speed of light, these high-energy particles can penetrate satellite shielding.

Characterizing the particles and their flux is crucial for astrophysics research and assessing potential damage to spaceborne biological tissue and man-made systems. It is also essential for developing countermeasures against this radiation~\cite{Workman:2022ynf}. When these particles approach Earth's magnetic field, their trajectories change, diminishing the particle flux at lower altitudes. Additionally, particles that reach Earth's atmosphere interact with it, and the majority of primary cosmic particles are absorbed before reaching earth's surface. Therefore, conducting in-situ spaceborne measurements is invaluable for understanding the characteristics of primary cosmic rays. Radiation-monitoring instruments, with various objectives and architectures, have been launched for space measurements (e.g.~\cite{BOEZIO2020103765}).

Radiation damage to spaceborne systems is categorized into cumulative damage and single-event effects (SEE)~\cite{SMAD} ~\cite{Katz2021}. 

Cumulative damage arises from numerous particle-matter interactions, primarily caused by electrons, protons, and heavier ions of the primary cosmic ray spectrum, as well as electrons and protons trapped in the Van Allen belts~\cite{VanAllen}. Cumulative damage effects involve multiple particles incrementally disturbing the lattice structure until the electronic device ceases to function properly. These effects encompass total ionizing dose (TID) and non-ionizing energy loss (NIEL), also known as displacement damage (DD). Damage to optical components is mainly DD-related, caused by alterations to the lattice structure due to nuclear displacement of lattice atoms by incident particles.

SEEs result from a single particle initiating soft errors, that can be removed or fixed, or catastrophic damage to electronic components. Soft errors include single event upsets (SEU), such as bit-flips in memory cells. Permanent damage may result from hard/catastrophic single-event effects (CSEE), occurring when a traversing particle triggers a parasitic device within the semiconductor, causing an internal short-circuit. Consequently, due to thermal dissipation, high currents quickly damage the semiconductor or the component's internal conductors.

This paper provides a comprehensive overview of the design and development of an innovative spaceborne cosmic ray detector system, named the COTS-Capsule. The COTS-Capsule was successfully deployed to the International Space Station (ISS) in 2021. Through in-situ measurements, the system gathered valuable data on cosmic radiation within the ISS environment. Serving as a precursor, this detector system lays the foundation for the forthcoming development of a particle detection and mitigation apparatus specifically designed based on the insights derived from the cosmic ray data collected aboard the ISS.

\section{The Mission}

The COTS-Capsule aims to enhance our understanding of space radiation within the ISS and facilitate the design of a CSEE mitigation apparatus. An experimental payload named COTS-Capsule Nanolab (CCN) was launched to the ISS in December 2021 aboard SpaceX's Dragon cargo spacecraft, carried by the Falcon 9 launch vehicle as part of CRS-24~\cite{CRS-24}. It orbited Earth in a circular low earth orbit (LEO) at an altitude of 420 km with an inclination of 51.6 degrees. Post-docking, the ISS crew installed the CCN within the Columbus module as depicted in \figref{FM_Nanoracks_Testing}. 

Operating successfully in the space environment for 48 days, the CCN meticulously gathered pertinent data during its tenure onboard the ISS. Following the mission's conclusion, the payload was securely returned to Earth as part of the CRS-25 mission~\cite{CRS-25}, marking the completion of this scientific endeavor.

\section{System Design}

The ISS functioned as both the housing for the CCN and the provider of power and communication services, featuring intermittent communication sessions approximately every two days with the ground station. Moreover, the ISS supplied atmospheric air pressure and forced airflow for thermal management, orbital station-keeping, and attitude control, and offered partial protection from space radiation. Although the ISS's shielding reduced the flux of lower-energy particles, higher-energy particles could penetrate the shielding and reach the CCN payload.

\subsection{Nanode Mainframe: The International Space Station Host}
\begin{figure*}
	\centering
		\includegraphics[width=\textwidth]{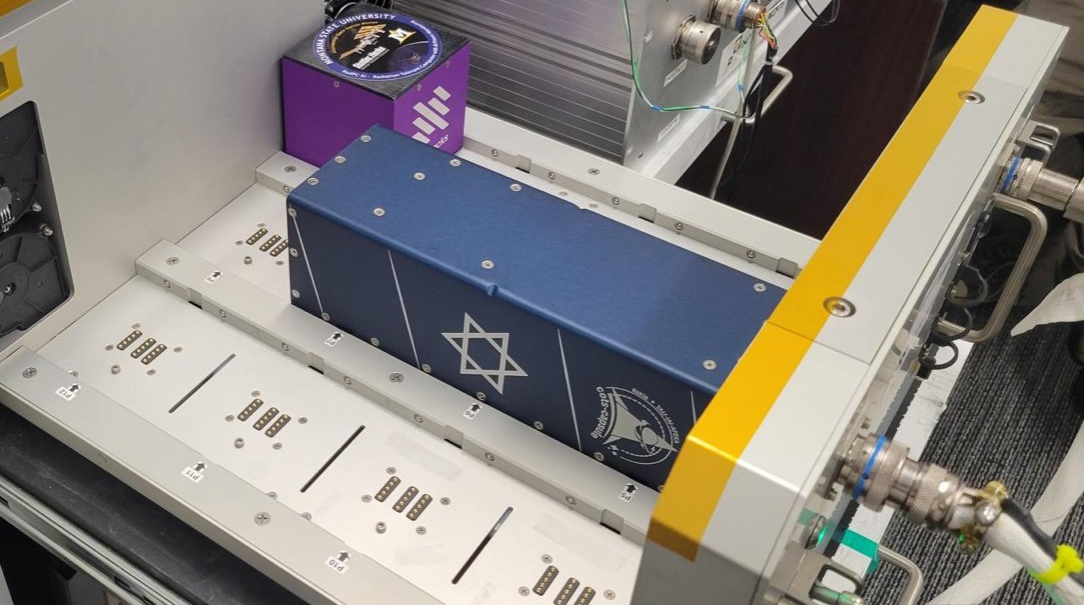}
	\caption{Installation and testing of the COTS-Capsule Nanolab (CCN, center in blue) flight module in the Nanoracks Nanode Mainframe (NNM) prior to launch. Image courtesy of NASA / Voyager Space.}
	\label{fig:FM_Nanoracks_Testing}
\end{figure*}

%
Voyager Space designed and installed the Nanoracks Nanode Mainframe (NNM) ~\cite{Nanode_ICD} within the ISS to serve as a versatile host capable of accommodating multiple Nanolab payloads of varying sizes. The NNM functions as the interface between Nanolab payloads, in our case the CCN, and the ISS, providing essential features such as mechanical mounting points, power supply, communication capabilities, and forced airflow thermal management. Furthermore, it extends various services to the Nanolabs, including memory storage and computing capabilities, as outlined in~\cite{Nanode_ICD}. \figref{FM_Nanoracks_Testing} depicts the installation of the CCN within the NNM testbench before launch.

The NNM is equipped with 12 independent slots, each featuring three communication and power connectors, facilitating the seamless integration of Nanolab payloads. A Nanode slot can accommodate a Nanolab payload measuring up to $200  \times 100  \times 100~\mathrm{mm}^3$ (similar to a 2U CubeSat form factor) in an upright position. Note that a Nanolab payload can span across multiple slots, and the NNM can accommodate multiple payloads simultaneously. However, the combined total capacity of these payloads should not exceed 24U.

\subsection{COTS-Capsule Nanolab: The Experiment Payload}

\begin{figure*}
	\centering
		\includegraphics[width=\textwidth]{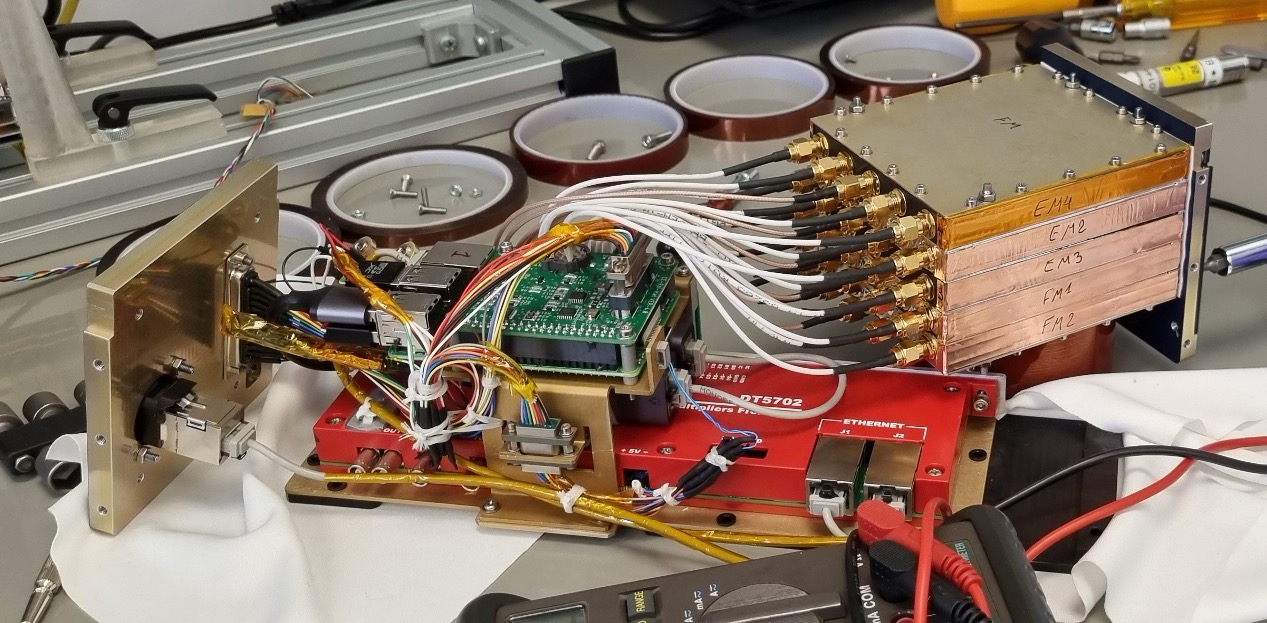}
	\caption{Open CCN Module during initial testing. Left: connector panel, right: particle hodoscope, bottom-center: DT5702, top-center: Raspberry Pi computer board, and peripheral electronics.}
	\label{fig:CCN_FM_Assembly}
\end{figure*}

The CCN payload, as illustrated in \figref{FM_Nanoracks_Testing} and  \figref{CCN_FM_Assembly}, conforms to a standard 3U Cubesat form factor with dimensions of $300 \times 100 \times 100~\mathrm{mm}^3$ and a mass of $2.5$ kg.

The scientific instrument includes five Scintillator Silicon Photo-Multiplier Particle Detectors (SSPDs) in a particle hodoscope configuration~\cite{SSPD} and its acquisition system based on CAEN's DT5702 module~\cite{DT5702}. 

The CCN also encompasses an electronics bay, housing a Raspberry Pi (RPi) 2B single-board computer~\cite{RPi}, data storage, communication adapters, wiring, and electronics required for the CCN's housekeeping and peripheral measurements. 

In the design of the CCN, considerations of system robustness and reduced vulnerability to space-induced failures, as well as other potential failures, were pivotal. The CCN is designed to operate autonomously, relying solely on an external power source without requiring communication, command, control, or data downlink. Upon receiving power, all systems initiate and become operational, allowing the CCN to gather and store data and telemetry. Regular self-tests are conducted, and the CCN incorporates failure-mitigating functions such as current monitoring, data-rate monitoring, and multiple independent and redundant watchdog mechanisms.

The CCN is enclosed within an aluminum enclosure. All instruments and electronics within the enclosure are galvanically insulated from the chassis except for specific bypass capacitors necessary for noise reduction. The enclosure serves multiple essential functions. Primarily, it encapsulates all components for operational purposes and facilitates safe handling by the astronaut crew. This includes encapsulating parts that might come loose in an unexpected failure event due to vibrations during launch and insulating the exterior from high voltages within the payload. Additionally, the enclosure serves to mitigate electromagnetic interference (EMI).

\subsection{COTS-Capsule Nanolab: Scientific Instrument Design}

\subsubsection{Hodoscope}

The particle hodoscope is composed of five SSPDs~\cite{SSPD}. These detectors provide precise measurements of the impinging position and energy deposition of incident particles. Engineered with a focus on functionality within the unique conditions of spaceflight in the ISS environment, the detector's design permits the stacking of multiple units. These detectors were stacked coaxially, sandwiched between additional aluminum cover plates, and secured with screws that traverse the entire apparatus \figref{CCN_FM_Assembly}. The resulting particle detector was mounted aboard the ISS, pointing toward Zenith.

The detectors within the hodoscope operate independently, individually receiving Silicon Photo Multiplier (SiPM) bias voltage from the CAEN DT5702 acquisition system, and returning analog signals from the SiPM sensors, which the acquisition system reads. The electronic connections are established using SMA-type connectors and RF coaxial cables, chosen to ensure signal integrity and minimize electronic interference that could potentially alter and impact the signals.

The particle hodoscope facilitates three-dimensional track reconstruction for impinging particles and offers an estimation of energy deposition per SSPD. Subsequently, the along-track linear energy transfer (LET) profile can be estimated which allows characterizing an incident particle by fitting it with the closest Bethe-Bloch curve~\cite{Workman:2022ynf}. 

The full characterization of an impinging particle depends on its track within the hodoscope and whether it has deposited all its energy within the detector or departed with residual energy. In addition to single-particle event characterization, the hodoscope provides a macro characterization of the particle field by evaluating particle fluence and angular profile.

\begin{figure*}
	\centering
     \begin{tabular}{cc}
      \includegraphics[width=0.48\textwidth,height=233pt]{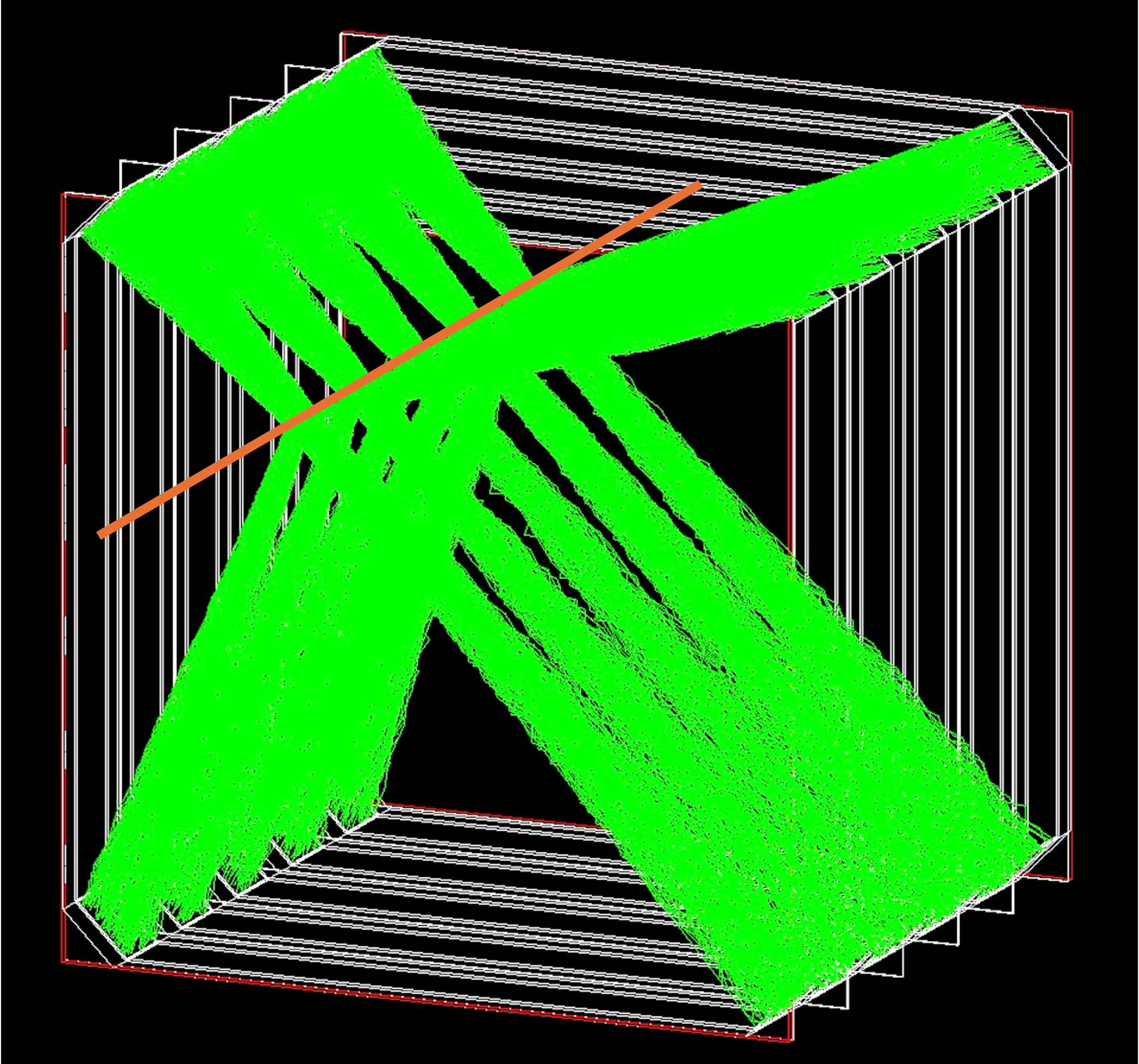} & \includegraphics[width=0.48\textwidth]{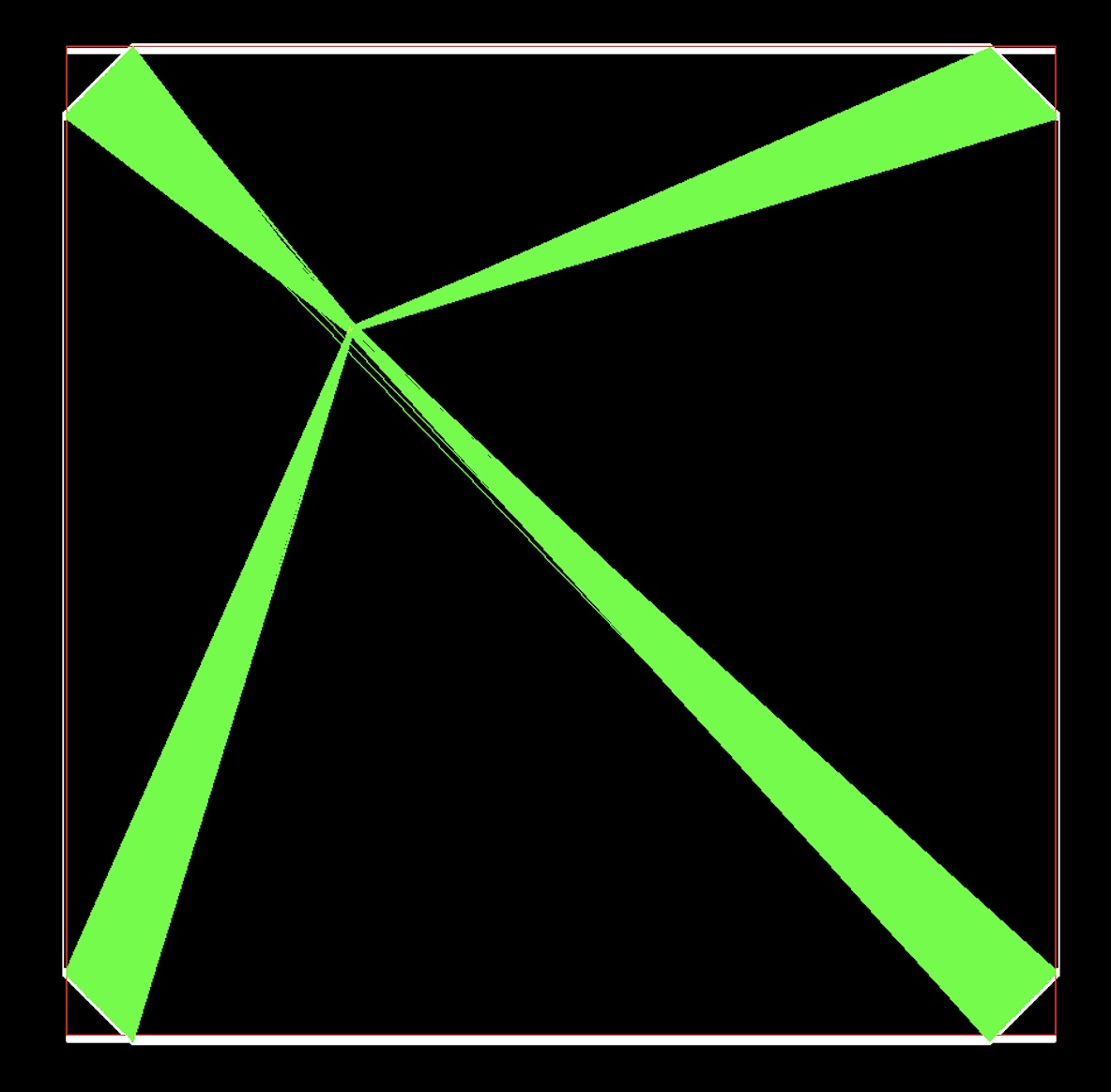}\\
      (a) & (b)\\
    \end{tabular}
    \caption{Particle trajectory (depicted in orange) of a 2~GeV Muon passing through a five-scintillator hodoscope (depicted in white). The simulations show photons (depicted in green) emitted when the muon interacts with the scintillator material. The photons shown are only those that ultimately strike one of the SiPM sensors located at the truncated corners of the scintillators. (a) provides a three-dimensional view, while (b) offers a forward-facing perspective.}
  \label{fig:hodoscope_simulations}
\end{figure*}

The GEANT4 simulation toolkit~\cite{GEANT4:2002zbu} was employed to conduct simulations of the SSPD and the hodoscope. A depiction of a 2~GeV Muon traversing the hodoscope is provided in \figref{hodoscope_simulations}. As the particle traverses the hodoscope perpendicularly, photons are emitted in all directions along its path within the scintillator material. The visualization shows only those photons that were detected by one of the SiPM sensors located on the truncated corner faces of the prism-shaped scintillators~\cite{SSPD}. 

\subsubsection{Acquisition System}

The data acquisition system employed in the CCN is based on CAEN's DT5702 module, incorporating Weeroc's Citiroc-1A readout Application Specific Integrated Circuit (ASIC) as its core component~\cite{fleury2014petiroc}. To our knowledge, this was the first time the CITIROC-1A ASIC was employed in a space-borne experiment.

The DT5702 module is powered by the NNM's 5V power supply, distributed within the CCN by its power electronics distribution unit. The data retrieved from the DT5702 is transmitted via Ethernet to the CCN's RPi on-board computer, which in turn stores it across redundant micro-SD memory cards. 

The DT5702 detects and measures signals from the twenty SiPM sensors distributed among five SSPDs. Connectivity to the detectors is facilitated through twenty-five RF cables soldered to a coax breakout Printed Circuit Board (PCB), as depicted in \figref{Electronic_diagram}. Additionally, the DT5702 supplies an approximate 28V bias voltage to the detectors. Onboard remote configuration capabilities are possible, allowing for the adjustment of bias voltage and other parameters mid-flight.

The original design of the DT5702 module falls short of utilizing the complete dynamic range achievable with the CITIROC-1A ASIC, restricting its functionality to a "high-gain" configuration. This limited the dynamic range available from the ASIC allowing detection and characterization of only low LET cosmic particles. To effectively capture the full dynamic range and characterize both high and low LET particles, a modification to the DT5702 was necessary.

This modification was performed with the support of CAEN and Bern University. By modifying the firmware we successfully introduced a "low-gain" configuration to the DT5702 alongside the existing "high-gain" setup. This modification resulted in an eightfold increase in the module's dynamic range. Following rigorous testing in the laboratory, the updated firmware was successfully deployed and operated in space, demonstrating its capacity to accommodate the broader dynamic range.

To insulate the DT5702 module from the chassis, it is mounted on an aluminum interface plate secured to a POM-C insulating frame and affixed to the top cover of the CCN as depicted in \figref{CCN_FM_Assembly}. The heat dissipation mechanism involves the use of Taica $\alpha$GEL\textsuperscript{\texttrademark} thermal pads. These pads fill the gap between the frame and the cover, facilitating heat conduction to the exterior case which is cooled onboard the ISS by forced airflow.

\subsection{COTS-Capsule Nanolab: Electronic Design}

%
\begin{figure*}
	\centering
		\includegraphics[width=\textwidth]{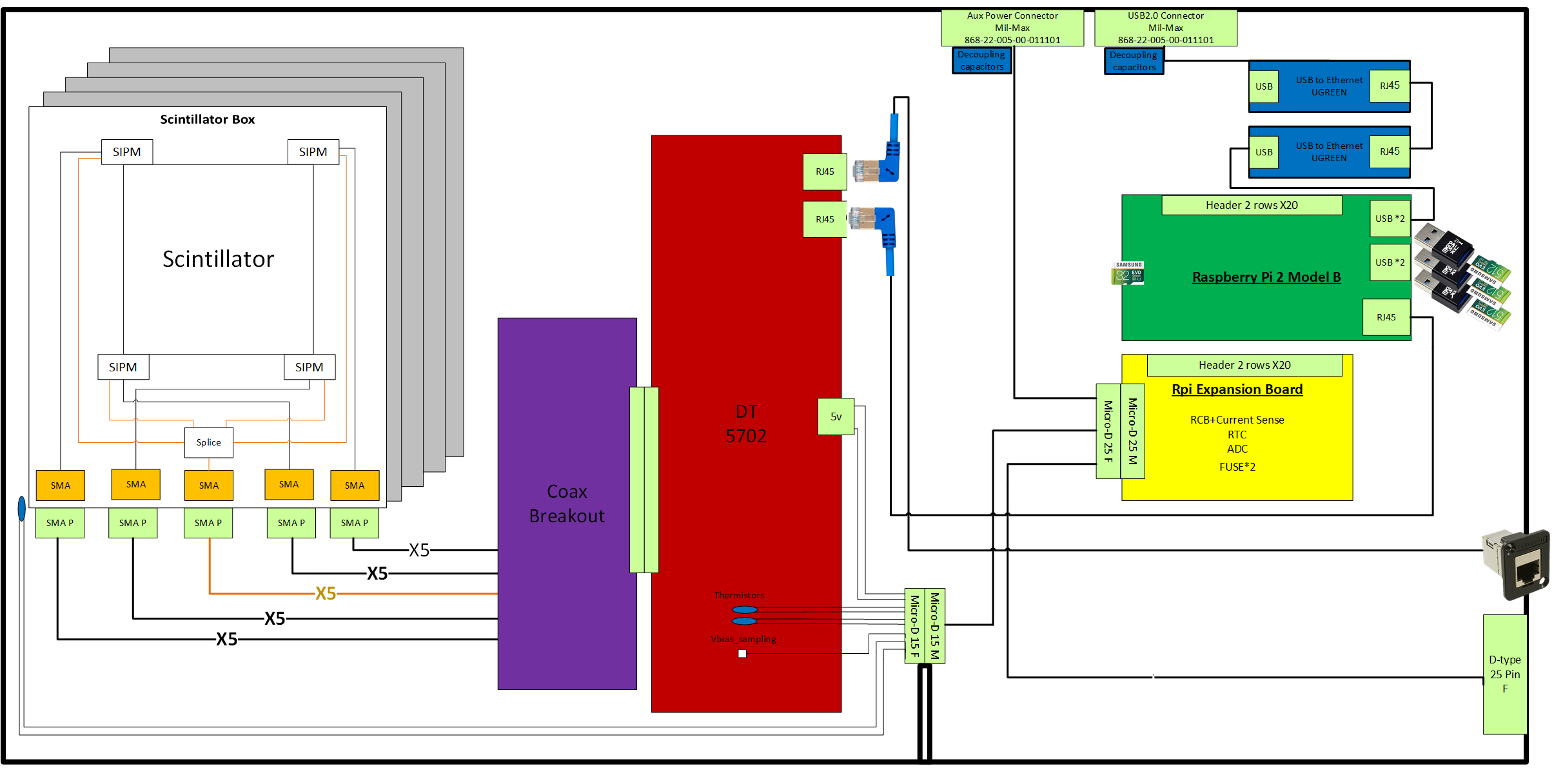}
	\caption{CCN: electronic system schematics. Top - two Mill-Max-type power and communication connectors towards the Nanoracks Nanode Mainframe and ISS. Left - particle hodoscope with five stacked SSPDs. Center - DT5702 acquisition module and RF-coax PCB adapter connecting the DT5702 to the hodoscope. Right - Electronics bay with an RPi single-board computer, the custom shield board, two communication adapters, and several micro-SD storage cards and adapters, and two test connectors.}
	\label{fig:Electronic_diagram}
\end{figure*}

A schematic diagram of the CCN electronic system is provided in \figref{Electronic_diagram}. 

The CCN interfaces with the NNM using two Mill-Max-type connectors, where one provides the 5V power supply, and the other facilitates a USB connection. Additionally, the CCN features two test connectors exclusively employed for lab testing. These connection options encompass an RJ-45 connector for Ethernet communications with the onboard electronics and a 25-pin D-type connector with various test functions.

It is essential to acknowledge that not all components underwent explicit design or testing for resilience to space radiation. The decision to utilize the RPi single-board computer, known for its successful use on the ISS, was motivated by a priority on reliability. However, the inherent variability among RPi family models, produced by different manufacturers with distinct component lots, introduced uncertainties regarding potential failure risks~\cite{Guertin7242974049}. The DT5702 acquisition module and the communication adapters were neither designed nor tested to space radiation.

The custom shield board, designed for the CCN's RPi, primarily integrated space-qualified electronics. In addition, it plays a role in addressing radiation-induced faults in other CCN modules. This is implemented by a current-limit-and-disconnect module which protects the DT5702 in the event of detectable radiation-induced over-currents. This protection circuit autonomously power-cycles the acquisition system multiple times before ceasing operations and requesting guidance from ground control. In addition, the custom shield board monitors and maintains autonomous operation in the challenging, no-maintenance, and radiation-prone space environment. Its functionalities also include a real-time clock, as well as temperature, voltage, and current monitoring.

The Samsung EVO 32GB MB-MP32G microSD card ~\cite{samsung-microsd_32GB} is employed for storing and running the operating system on the RPi. This choice was made due to its known compatibility with the RPi, improved reliability, and its resistance to X-ray radiation. The latter characteristic is noteworthy as it suggests a potential suitability for withstanding space-based SEUs.

For data storage in space, three redundant micro-SD cards were used, including two SAMSUNG EVO Plus 512GB - MB-MC512H~\cite{samsung-microsd-512gb} and one Samsung EVO Select 512GB - MB-ME512H~\cite{samsung-evo-select-512gb}. These cards were connected to the RPi's USB ports through three micro-SD to USB adapters. We deliberately diversified both the microSD cards and their adapters to increase system redundancy, considering the unknown susceptibility to radiation. Additionally, we configured the memory cards to have internal redundancy using RAID1 and the Btrfs file system, as described in detail in \ref{software_design}.

\subsection{COTS-Capsule: Software Design} \label{software_design}

The onboard Raspberry Pi (RPi) computer operates with Raspberry Pi OS version 10 (code-named "Buster")~\cite{raspberrypios}. The software is designed for fully autonomous operation in the space environment and includes several key modules. The primary module is dedicated to data collection. Additional modules are responsible for system self-tests, diagnostics, and telemetry gathering. Telemetry can then be downlinked for in-depth analysis on the ground. Other modules manage system maintenance tasks, such as data backups, disk space management, and error detection and correction. The main functions are detailed below:

\subsubsection{Autonomy and Reliability}

To ensure self-recovery from unpredicted failures, several software watchdogs are implemented.

Several measures are implemented in the software to enhance the CCN's data resilience against radiation-induced errors, encompassing both soft errors like SEUs and catastrophic errors such as latch-up. The RPi's OS is installed on the Btrfs file system in DUP mode to mitigate data corruption risks. Btrfs, a contemporary Copy-on-Write (COW) file system, offers various advantageous features such as checksums, automatic error correction, and transparent compression.~\cite{btrfs}

As previously mentioned, the three external storage micro-SD cards are configured in RAID1 mode with the Btrfs file system. In addition, the RPi's built-in system watchdog is activated to automatically reboot the system if it becomes unresponsive.

\subsubsection{Data collection service}
%

To interact with the DT5702 acquisition module, a custom systemd service is implemented. This service communicates with the DT5702 using a custom data-link protocol.
Each event read from the DT5702 is stored as a separate line in the data file. Every line consists of a timestamp and forty simultaneous measurements acquired from all twenty SiPMs in two gain configurations: a "Low Gain" amplification configuration, and a "High Gain" amplification configuration. 

\subsubsection{Real-Time Monitoring}
%
The RPi and its custom shield board periodically monitor various temperature, current, and voltage measurements, logging them to a telemetry file through a dedicated custom service that initializes on boot.

If any sampled value deviates from the predefined nominal range, the service responds with either a warning or, depending on the specific value, powers down the DT5702 acquisition module. If the DT5702 is powered down, data collection ceases, and the data-collection service's watchdog automatically reboots the system after two hours of inactivity.

\subsubsection{Automatic tasks}
%
The system incorporates several scheduled tasks for maintenance purposes:
\begin{enumerate}
\item Disk Space Monitoring and Handling:
A daily scheduled task monitors free disk space on each partition of the internal micro-SD boot card. Under normal working conditions, a 'low disk space' situation should not occur. On top of this, old files are removed to guarantee sufficient disk space in case of abnormal conditions. 

\item Diagnostics and Error Correction:
A daily automatic task executes a file-system error check and correction to ensure the ongoing integrity of the file system.

\item Backup and Validation:
The boot partition of the RPi cannot utilize features of advanced filesystems. Thus, the partition is backed up and integrity checks are performed daily. 

\end{enumerate}

\section{Qualification and Test Campaign}

Before its launch, the CCN underwent a "qualification campaign" to ensure reliable operation in the ISS environment. Testing was also conducted to ensure compliance with safety criteria for the launcher, ISS, and crew.

Every component and sub-component of the CCN, including the hodoscope SSPDs, electronic boards, and Commercial-Off-The-Shelf (COTS) modules and components, underwent individual testing. Subsequently, integration tests were carried out to assess seamless interaction among all the components. Throughout the space qualification campaign, the CCN underwent an extensive series of tests focused on validating and verifying the system's functionality, integrity, robustness, and reliability to ensure its survival through launch and correct operation aboard the ISS.

The qualification campaign's test sequence focused on functional and comparative testing of several manufactured CCN modules. This encompassed continuous monitoring of module behavior throughout the qualification campaign, during, and after each stress test.

The testing phase for the modules included a comprehensive suite of functionality tests to ensure robust performance under various conditions. These included characterization of boot times, power, and current consumption across both nominal and extreme operational states. 
The tests were conducted at voltage levels $\pm10 \%$ around nominal values to assess the system's resilience against power instabilities. 

Temperature testing was specifically tailored to the CCN’s operational environment aboard the ISS, with a controlled range from $+15$ \degree C to $+40$ \degree C. 

The rigorous testing regimen culminated in long-duration functionality tests that evaluated software behavior and the correct operation of watchdog mechanisms. Special attention was given to ensuring long-term autonomous and maintenance-free operation in simulated environments, such as the NNM simulator. Throughout these tests, various parameters of the CCN were measured both internally by the CCN as well as by external laboratory equipment. These tests served several purposes: qualifying the modules for spaceflight, evaluating the CCN's performance, and calibrating the scientific instrument with secondary cosmic particles. 

Additionally, safety and interface requirements testing were conducted to comply with the NNM requirements, including chassis grounding and bonding tests, as specified by~\cite{Nanode_ICD}. This comprehensive evaluation was further bolstered by a detailed safety review conducted by NASA and a thorough examination by Voyager Space.

\subsection{Vibration Testing}
%
The purpose of the vibration testing was to validate the ability of the primary mechanical structure and all electronic and mechanical components to withstand the vibrations experienced during launch.

Random vibration testing was conducted following the guidelines outlined in~\cite{Nanode_ICD}. The stress tests were performed on both the Proto-Flight Module (PFM) and the Flight Module (FM), employing the soft-stow profile. This profile represents the worst-case vibration scenario expected during launch. The soft-stow profile entails a $60$ second vibration test per axis at $5.76$ G-rms. The CCN was affixed to a DONGLING ES-1-150 Air-Cooled Shaker, simulating conditions similar to those anticipated during stowage onboard SpaceX's Dragon capsule.

An extreme vibration testing profile was applied to a single SSPD, utilizing the "hard-stow" profile~\cite{Nanode_ICD}, as this component was deemed the most sensitive within the CCN. The detector successfully passed these tests.

Following all vibration tests, the CCN demonstrated both correct performance and mechanical integrity.

\subsection{Integration Testing Within the Nanode Mainframe}
%
Several integration tests were conducted within the NNM simulator at Voyager Space's facility. Initial integration tests were performed with the Mock-Up module. The objective was to validate the mechanical, electrical, and communications aspects of the CCN, as designed according to the NNM ICD~\cite{Nanode_ICD}. Notably, this test excluded testing the hodoscope, as it was not incorporated in the Mock-Up.

Following the successful environmental and Mock-Up tests with the NNM simulator, the PFM and FM underwent pre-flight integration with the NNM simulator. During integration, an anomaly was identified where a high rate of false events was detected by the particle hodoscope detector. An electrical investigation uncovered an unexpected $8~$MHz, $0.5~$$V_{ptp}$ electronic interference emanating from within the NNM simulator's chassis.

This unexpected interference propagated through capacitive coupling from the NNM simulator's chassis to the CCN's ground, resulting in hodoscope false triggers. Given that the NNM has already been installed onboard the ISS, revisions were performed to both the PFM and FM modules to mitigate the interference within the CCN. After simulations and testing, $12~$nF filtering capacitors were added between the chassis and all of the CCN's power and ground inputs as depicted in \figref{Electronic_diagram}. The interference to the PFM was successfully eliminated. However, while the noise coupled to the FM was significantly suppressed, some residual noise remained detectable. Since this noise was parasitic in nature and its magnitude in the space environment was unknown we opted to launch the PFM module to the ISS instead of the FM module.

\subsection{Hodoscope Calibration}

\subsubsection{Laboratory Hodoscope Calibration}
%
Upon completing the assembly and comprehensive testing of the CCN, calibration of its hodoscope became possible. This crucial step, addressing manufacturing tolerances in mechanical and optical aspects, was essential to produce precise particle measurements. Calibration was pivotal in estimating energy deposition, triangulating particle impact locations, and achieving accurate particle track estimations. The process in the lab utilized measurements of secondary cosmic particles (muons) as outlined in~\cite{SSPD}.
%

\subsubsection{In Space Hodoscope Calibration}

After the launch of the PFM CCN module to the ISS, recalibration was conducted in situ using primary cosmic particles, predominantly protons and heavy ions. This step was essential to verify that the launch to the ISS did not induce mechanical or optical changes in the particle hodoscope and to ensure precise measurements in space. The calibration procedure followed the process outlined in ~\cite{SSPD}, which was implemented in the laboratory before the launch.

Two fundamental assumptions underpinned the calibration approach:

\begin{enumerate}
\item The flux of cosmic rays in space is largely independent of rotation around the Zenith axis.
\item Heavy ions traversing the entire length of the hodoscope do so in approximately a straight line.
\end{enumerate}

For calibration, we considered only particle events where the particle traversed the entire hodoscope, receiving adequate signals in all of the hodoscope's twenty SiPM sensors. Data from both the high-gain and low-gain dynamic ranges were combined. This approach allowed the calibration of the hodoscope with particle events that produced a mix of low and high signals across the twenty SiPM sensors, thereby increasing the number of particle events used in the calibration process. Calibration data was collected over 48 days, resulting in a $1.1 \times 10^7$ particle events dataset.

Following the calibration procedure, we compared the calculated SiPM constants ($c_{ij}^{(k)}$) ~\cite{SSPD} pre-launch (in the lab) and post-launch (in space). Space calibration resulted in variations of up to $15\%$ between the twenty SiPM calibration constants. When comparing the constants derived from lab measurements to those obtained from space measurements, there was less than a $2\%$ change per SiPM constant. The results indicate no discernible changes in the hodoscope's performance before and after launch, suggesting that the CCN survived the launch intact and without changes.

Each impinging particle's track can be estimated using the \textit{Non-perpendicularly Impinging Particle Position Estimation Algorithm} (Algorithm $2$) \cite{SSPD}. This involves analyzing the SSPDs in the hodoscope and fitting the lowest RMS straight track to the derived impinging position estimates. By comparing the derived estimated particle impinging positions on a specific SSPD to the estimated tracks, we can determine the position estimation error for each impinging particle.

\begin{figure}[htbp]
\centering
\includegraphics[width=0.7\textwidth]{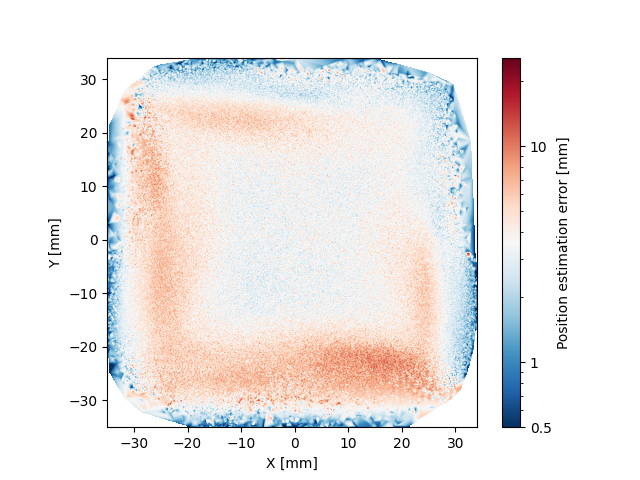}
\caption{Heat map illustrating the estimation errors of impinging particle positions according to the Non-Perpendicularly Impinging Particle Algorithm (Algorithm $2$). The dataset comprises $1.1 \times 10^7$ in-situ particle events. The estimation uncertainties of impinging particle positions have a standard deviation of $3.7~$mm.}
\label{fig:heat_map_position_estimation_errors_space}
\end{figure}

A heat map illustrating the estimation errors of impinging particle positions, according to Algorithm $2$, is shown in \figref{heat_map_position_estimation_errors_space}. The standard deviation obtained for the entire sensitive area of the SSPD is $3.7~$mm. This heat map closely resembles the features and values of the heat map produced in \cite{SSPD} by simulating the SSPDs in GEANT4 and applying Algorithm $2$. Minor differences between the experimental results and theoretical simulations are likely due to manufacturing variations.

\begin{figure*}
	\centering
		\includegraphics[width=\textwidth]{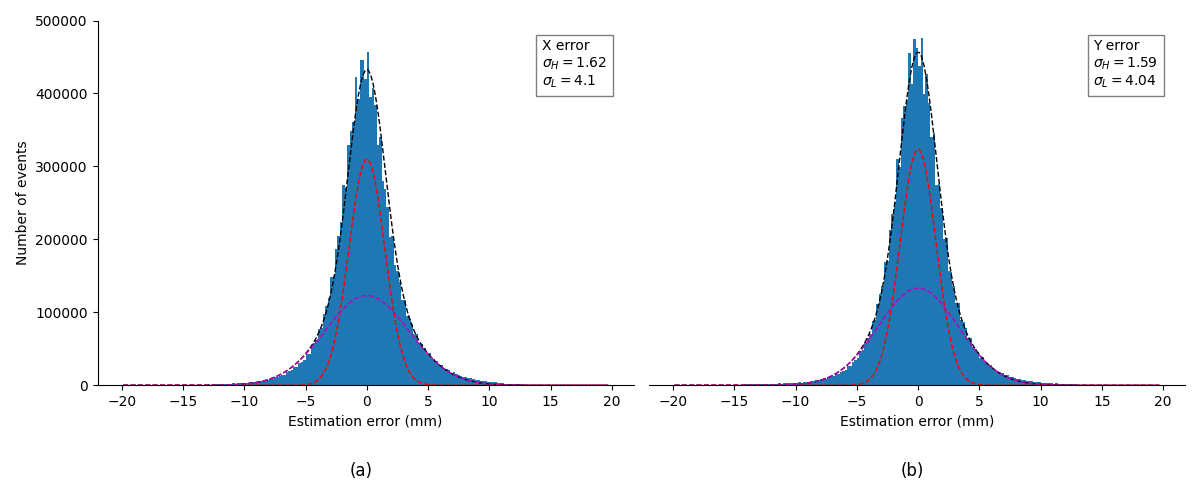}\label{fig:gaussians}
    \caption{Histograms showing the distribution of estimation errors for impinging particle positions obtained using the Non-Perpendicularly Impinging Particle Algorithm (algorithm 2). (a) displays the histogram of errors in the x-coordinate estimation, while (b) shows errors in the orthogonal y-coordinate estimation. The dataset comprises $1.1 \times 10^7$ in-situ measurements. The dashed black line represents a two-Gaussian sum fit to the data. The fit is the sum of the red dashed line Gaussian fit and the purple dashed line Gaussian fit.} 
  \label{fig:gaussian_fit_space}
\end{figure*}

\figref{gaussian_fit_space} depicts the distribution of estimation uncertainties for impinging particle positions according to the Non-Perpendicularly Impinging Particle Algorithm. The dataset comprises $1.1 \times 10^7$ in-situ particle events. The dashed black line represents a two-Gaussian sum fit to the data, consisting of the red dashed line Gaussian fit and the purple dashed line Gaussian fit.

\section{Summary and Conclusions}
We developed, simulated, manufactured, tested, launched, and calibrated the CCN payload aboard the ISS.

The CCN comprises a scientific instrument featuring a hodoscope and acquisition electronics, an electronics bay, and interfaces with both the NNM and the ISS. This experiment marks the first successful use of SSPDs and CITIROC-1A analog readout ASICs in space. 

The hodoscope comprises five SSPDs arranged vertically in a particle telescope configuration. These detectors enable the detection of both primary and secondary cosmic particles in laboratory settings and in space. The hodoscope accurately estimates impinging particle positions and measures energy deposition in each SSPD. Additionally, it facilitates track reconstruction for each particle event. In instances where a particle stops within the hodoscope, it also estimates the initial particle energy and identifies its type.

The payload detected and recorded detailed cosmic particle events in the unique ISS environment, including timestamped data on particle position and energy deposition per detector for each event. It utilized an extended dynamic range, allowing the detection of both high-LET and low-LET particles.

A thorough test campaign systematically validated and verified the experiment's design, identifying and addressing several deficiencies before launch. Hodoscope calibration was conducted in the laboratory and in space using secondary and primary cosmic particles. The calibration process yielded consistent results, with deviations of up to $2\%$ between pre-launch and post-launch, confirming reliable operation in the space environment without any launch-related failures. Furthermore, the hodoscope measurements in space validated and verified the results of the GEANT4 simulations.

After calibration with primary cosmic particles in space, the impinging particle position estimation error in each SSPD was shown to be $3.7~$mm, significantly better than the particle position estimation error obtained in the lab with secondary cosmic rays ($5.1~$mm) \cite{SSPD}. This improvement is attributed to the higher average LET of primary cosmic particles detected in space compared to secondary cosmic particles in the laboratory. The higher average LET increases the signal-to-noise ratio in the SSPD's SiPMs, thereby improving the accuracy of position and deposited energy estimations per SSPD.

\section{Appendix - Raspbbery Pi 2B hardware bug}
%
During laboratory testing, the RPi computer experienced random reboots, seemingly linked to the combination of changing CPU operating frequency and elevated ambient temperature.

The default CPU governor in RPi OS is a scaling governor, adjusting the CPU frequency to 600MHz when the OS is idle and scaling up to 900MHz when required.

Upon careful examination, it was found that frequent up and downscaling of the CPU frequency, particularly when the ambient temperature exceeded $50\degree$ Celsius, triggered a hardware reset in the RPi. This issue was replicated across multiple RPi boards, including scenarios with no peripheral devices and with various Linux and FreeBSD OS versions.

A workaround involved fixing the CPU frequency to 600MHz and disabling frequency scaling, effectively mitigating the reboot problem without noticeable performance loss.

\section{Acknowledgements}

We extend our heartfelt gratitude to all individuals and organizations whose contributions were instrumental to the success of the COTS-Capsule mission to the International Space Station. Special thanks to Eitan Stibbe and the Ramon Foundation for their leadership in initiating and overseeing the realization of the Rakia mission. The ISS deployment opportunity was made available by Voyager Space through its Space Act Agreement with NASA's U.S National Lab. Eljen, for their exceptional craftsmanship, dedication, and meticulous attention to detail. CAEN, Bern University, and Weeroc for their invaluable support with the DT5702 and CITIROC-1A. The Soreq Nuclear Research Center provided essential guidance and technical assistance through their space environment and machine shop teams. Our colleagues at Tel Aviv University, including the Nano-Satellite lab, and the Physics Machine Shop, for their guidance and support throughout the project. Special thanks to Meny Raviv Moshe, Dr. Igor Zolkin and Prof. Yan Benhammou. Israel Aerospace Industries for their expertise, insightful feedback, and constant support. Etzion acknowledges the support of the Canada First Research Excellence
Fund. Lastly, our sincere appreciation goes to Tal Ahitov from Samsung.

\bibliographystyle{unsrt}
\bibliography{cas-refs}

\begin{thebibliography}{10}

\bibitem[Workman et~al.(2022)]{Workman:2022ynf}
R.~L. Workman et~al.
\newblock Review of particle physics.
\newblock {\em PTEP}, 2022:083C01, 2022.

\bibitem[Boezio et~al.(2020)]{BOEZIO2020103765}
M.~Boezio, R.~Munini, and P.~Picozza.
\newblock Cosmic ray detection in space.
\newblock {\em Progress in Particle and Nuclear Physics}, 112:103765, 2020.

\bibitem[Wertz et~al.(1999)]{SMAD}
J.~R. Wertz, W.~J. Larson, D.~Kirkpatrick, and D.~Klungle.
\newblock {\em Space Mission Analysis and Design}, volume~8.
\newblock Springer, 1999.

\bibitem[Katz et~al.(2021)]{Katz2021}
S.~Katz, U.~Goldvais, and C.~Price.
\newblock The connection between space weather and single event upsets in polar low earth orbit satellites.
\newblock {\em Advances in Space Research}, 67:3237--3249, 2021.

\bibitem[Van Allen(1958)]{VanAllen}
J.~A. Van Allen.
\newblock Doughnuts of radiation ring earth in space.
\newblock Victoria Advocate (Texas), Associated Press, December 28, 1958, p. 1A, 1958.

\bibitem[SpaceX(2020a)]{CRS-24}
SpaceX.
\newblock CRS-24 mission.
\newblock \url{https://www.spacex.com/launches/crs-24-mission/}, 2020.

\bibitem[SpaceX(2020b)]{CRS-25}
SpaceX.
\newblock CRS-25 mission.
\newblock \url{https://www.spacex.com/launches/crs-25/}, 2020.

\bibitem[Nanoracks LLC(2020)]{Nanode_ICD}
Nanoracks LLC.
\newblock Interface definition document (IDD), Nanoracks Mainframe (Nanode), NR-NANODE-S00001.
\newblock pages 1--38, 2020.

\bibitem[Simhony et~al.(2024)]{SSPD}
Y.~Simhony, Y.~Orlov, D.~Bashi, A.~Segal, O.~Amrani, and E.~Etzion.
\newblock Scintillator-SiPM detector for tracking and energy deposition measurements.
\newblock {\em arXiv:2406.19652 [physics.ins-det]}, 2024.

\bibitem[CAEN(2020)]{DT5702}
CAEN.
\newblock 32 channel SiPM readout board for cosmic rays veto boxed.
\newblock \url{https://www.caen.it/products/dt5702/}, 2020.

\bibitem[Raspberry Pi Foundation(2020)]{RPi}
Raspberry Pi Foundation.
\newblock Raspberry Pi.
\newblock \url{https://www.raspberrypi.com/}, 2020.

\bibitem[Agostinelli et~al.(2003)]{GEANT4:2002zbu}
S.~Agostinelli et~al.
\newblock GEANT4--a simulation toolkit.
\newblock {\em Nucl. Instrum. Meth. A}, 506:250--303, 2003.

\bibitem[Fleury et~al.(2014)]{fleury2014petiroc}
J.~Fleury, S.~Callier, C.~de~La~Taille, N.~Seguin, D.~Thienpont, F.~Dulucq, S.~Ahmad, and G.~Martin.
\newblock Petiroc and Citiroc: Front-end ASICs for SiPM read-out and ToF applications.
\newblock {\em Journal of Instrumentation}, 9(01):C01049, 2014.

\bibitem[Guertin(2021)]{Guertin7242974049}
S.~M. Guertin.
\newblock Raspberry Pis for space guideline, 2021.

\bibitem[Samsung(2020a)]{samsung-microsd_32GB}
Samsung.
\newblock Samsung EVO microSDHC 32GB.
\newblock \url{https://www.samsung.com/us/business/support/owners/product/evo-microsdhc-32gb/}, 2020.

\bibitem[Samsung(2020b)]{samsung-microsd-512gb}
Samsung.
\newblock Samsung EVO Plus microSD card 512GB.
\newblock \url{https://www.samsung.com/uk/memory-storage/memory-card/evo-plus-microsd-card-512gb-mb-mc512ha-eu/#specs}, 2020.

\bibitem[Samsung(2020c)]{samsung-evo-select-512gb}
Samsung.
\newblock Samsung EVO Select microSDXC memory card 512GB.
\newblock \url{https://www.samsung.com/us/computing/memory-storage/memory-cards/evo-select-microsdxc-memory-card-512gb-mb-me512ha-am/#specs}, 2020.

\bibitem[Raspberry Pi OS(2020)]{raspberrypios}
Raspberry Pi OS.
\newblock \url{https://www.raspberrypi.com/documentation/}, 2020.

\bibitem[Btrfs(2020)]{btrfs}
Btrfs kernel documentation.
\newblock \url{https://docs.kernel.org/filesystems/btrfs.html}, 2020.

\end{thebibliography}

\end{document}